\documentclass[11pt]{article}

\usepackage{amsmath,amssymb}
\usepackage{latexsym}
\usepackage{graphicx}
\usepackage{bm}

\newcommand{\R}{{\mathbb R}}

\newcommand{\bee}{\begin{equation*}}
\newcommand{\eee}{\end{equation*}}
\newcommand{\be}{\begin{equation}}
\newcommand{\ee}{\end{equation}}

\def\Image{\mathop{\rm Im}}

\def\diam{\mathop{\rm diam}}
\def\R{\mathbb{R}}

\begin{document}
Plenary talk at the Chaos 2009 International Conference.

In the book "Topics on Chaotic Systems: Selected Papers from Chaos 2009
International Conference" Editors C.Skiadas, I.Dimotikalis,
Char.Skiadas, World Sci. Publishing, 2011

A. G. Ramm, Scattering by many small inhomogeneities
and applications

\title{Scattering by many small inhomogeneities
and applications}

\author{A. G. Ramm$\dag$\footnotemark[1]
\\
\\
$\dag$Mathematics Department, Kansas State University,\\
Manhattan, KS 66506-2602, USA
}

\renewcommand{\thefootnote}{\fnsymbol{footnote}}
\footnotetext[1]{Email: ramm@math.ksu.edu}

\date{}
\maketitle

\begin{abstract}
Many-body quantum-mechanical  scattering problem is solved
asymptotically when the size of the scatterers (inhomogeneities)
tends to zero and their number tends to infinity.

 A method is given for calculation of the number of small
inhomogeneities per unit volume and their intensities such that
embedding of these inhomogeneities in a bounded region
results in creating a new system, described by a desired potential.
The governing equation for this system is a non-relativistic
Schr\"odinger's equation described by a desired potential.

Similar ideas were developed by the author for acoustic and
electromagnetic (EM) wave scattering problems.

\end{abstract}

\noindent
{\bf PACS} 43.20.+g, 62.40.+d, 78.20.-e.\\
{\bf MSC}  35J10; 45F05; 74J25, 81U10; 81U40; 82C22

{\bf Keywords:} wave scattering by small inhomogeneities; metamaterials;
nanotechnology; refraction coefficient; negative refraction.

\section{Introduction}
In a series of papers, cited in the references, and in the
monograph \cite{R476}
the author has developed wave scattering theory by many small bodies of
arbitrary shapes and, on this basis, proposed a method for creating materials
with some desired properties, in particular, with a desired
refraction coefficient. The goal of
this paper is to describe
a new method for creating materials with a desired refraction coefficient
and to compare it with the method proposed earlier.
The new method is applicable to the problems which 
deal with vector fields and tensorial refraction coefficients. 
Our presentation deals with the simpler case of scalar
fields and uses paper \cite{R564}.

Let us formulate the statement of the problem.  Let $D\subset
\mathbb{R}^3$ be a bounded domain filled in by a material with a known
refraction coefficient. The scalar wave scattering problem consists of
finding the solution to the Helmholtz equation:
\begin{equation}
\label{eq1} L_0u:=[\nabla^2+k^2n_0^2(x)]u=0 \quad \text{in}\quad
\mathbb{R}^3,\quad\Image n_0^2(x)\ge 0, \end{equation}

\begin{equation}
\label{eq2}
u=u_0+v, \quad u_0:=e^{ik\alpha\cdot x},
\end{equation}

\begin{equation} \label{eq3} v=A_0(\beta,\alpha)\frac{e^{ikr}}{r}+
o(\frac{1}{r}), \quad r:=|x|\to\infty,\quad \beta:=\frac{x}{r}.
\end{equation}
The function $n^2_0(x)$ is assumed bounded,
Riemann-integrable,
\begin{equation} \label{eq4} n_0^2(x)=1 \quad
\text{in}\quad D':=\mathbb{R}^3\backslash D, \quad \Image n_0^2(x)\ge 0.
\end{equation}
The function $A(\beta,\alpha)$ is called the scattering
amplitude.  The wavenumber $k=\frac{2\pi}{\lambda}$, where $\lambda$ is
the wavelength in $D'$, $\alpha\in S^2$ is the direction of the incident
plane wave $u_0$, $S^2$ is the unit sphere in $\mathbb{R}^3$, $\beta\in
S^2$ is the direction of the scattered wave.  The solution to problem
\eqref{eq1}-\eqref{eq3} is called the scattering solution.  It is
well-known (\cite{R190}) that this solution exists and is unique under our
assumptions. These assumptions are:  $n_0^2(x)$ is bounded:
$\sup_{x\in\mathbb{R}^3}|n_0^2(x)|\le n_0=const$, $\Image n_0^2(x)\ge 0$,
$n_0^2(x)=1$ in $D'$. Since the solution $u$ is unique, the corresponding 
scattering amplitude  
$A_0(\beta,\alpha)$ is determined uniquely
if $n_0^2(x)$ is given. We assume $k>0$ fixed and do not show the
dependence of $A_0(\beta,\alpha)$ on $k$. The operator $L_0$ at a fixed
$k>0$ can be considered as a Schr\"odinger operator
$L_0=\nabla^2+k^2-q_0(x)$, where $q_0(x):=k^2-k^2n_0^2(x)$.

{\bf  Problem RC}:

{\it We want to construct a material in $D$ with a
desired
refraction coefficient, that is, with a desired function $n^2(x)$, $\Image
n^2(x)\ge 0$.}
$$ $$
Here RC stands for refraction coefficient.

{\it Why is this problem of practical interest?}

We give two reasons.

{\it First}, creating a refraction coefficient such that the corresponding
material has {\it negative refraction} is of practical interest.

One says that
a material has negative refraction if the group
velocity in this material is directed
opposite to the phase velocity.

{\it Secondly}, It is  of practical interest to  create a refraction 
coefficient such that the
corresponding
scattering amplitude $A(\beta)=A(\beta,\alpha)$ at a fixed $\alpha\in S^2$
(and a fixed $k>0$)
approximates  an arbitrary given function $f(\beta)\in L^2(S^2)$
 with any desired accuracy $\epsilon>0$. 

This problem we call {\it the problem of creating material with a desired
wave-focusing property.}

In Section 2 we explain how to create a material with a desired
refraction coefficient. In Section 3 we
compare our recipe for creating material with
a desired refraction coefficient with the recipe given in \cite{R552}.

\section{Creating material with a desired refraction coefficient by
embedding small inhomogeneities into a given material}

The basic idea of our method is to embed in $D$ many small inhomogeneities
$q_m(x)$ in such a way that the resulting medium can be described by
a desired potential $q(x)$.

 Let us assume that $p_M(x)$ is a
real-valued compactly supported bounded  function, which is a sum
of small inhomogeneities:
$$p_M=\sum_{m=1}^M q_m(x),$$ 
where $q_m(x)$ vanishes outside the ball
$B_m:=\{x: |x-x_m|<a\}$ and  $q_m=A_m$ inside $B_m$,
$1\leq m \leq M$, $M=M(a)$.

{\it The question is:

Under what conditions the field $u_M$, which solves the
Schr\"odinger equation with the potential $p_M(x)$, has
a limit $u_e(x)$ as $a\to 0$, and this limit
$u_e(x)$ solves the Schr\"odinger equation with a
desired potential $q(x)$?
}

We give a complete answer to this question in Theorem 1  below.

The class of potentials $q$, that can be obtained by our
method, consists of bounded, compactly supported,
Riemann-integrable functions. It is known that the set
of Riemann-integrable functions is precisely the set of
almost everywhere continuous functions, that is, the set
of bounded functions  with the set of
discontinuities of Lebesgue  measure zero in $\R^3$.
These assumptions on $q$ are not repeated
but are always valid when we write  "an arbitrary potential".

In fact, a more general set of potentials
can be constructed by our method. It is
mentioned below that for some class of unbounded potentials,
having  local singularities, which are absolutely integrable, our
theory remains valid.

Our result is as follows:

{\it Assume that $q(x)$ is an arbitrary Riemann-integrable in $D$
potential, vanishing outside $D$, where $D$ is an
arbitrary large but finite domain, and the functions $A(x)$ and
$N(x)$, which we can choose as we wish, are chosen so that 
$A(x_m)=A_m$ and $A(x)N(x)=q(x)$, where $N(x)\geq 0$.

Then the limit $u_e(x)$ of $u_M(x)$ as $a\to 0$ does exist, and solves
the scattering problem  with the desired refraction coefficient
$n^2(x)$:
\begin{equation}
\label{eq1'}
\nabla^2 u_e+k^2n^2(x) u_e:=\nabla^2 u_e +k^2 u_e-q(x) u_e=0,
\end{equation}
\begin{equation}\label{eq2'}
u=u_0+v,
\end{equation}
where $u_0$ solves problem  \eqref{eq1'} and $v$ satisfies the radiation
condition. }

The notation $u_e(x)$ stands for the effective field, which is the
limiting field in the medium as $M\to \infty$, or, equivalently,
$a\to 0$. Under our assumptions (see Lemma 1 below) one has
$M=O(\frac 1 {a^3})$.

The field $u_M$ is the unique solution to the integral equation:
\be\label{e3}
u_M(x)=u_0(x)-\sum_{m=1}^M \int_D g(x,y,k)q_m(y)u_M(y)dy,
\qquad g(x,y,k)=\frac {e^{ik|x-y|}}{4\pi |x-y|},
\ee
where $u_0(x)$ is the incident field, which one may take as
the plane wave, for example, $u_0=e^{ik\alpha \cdot x}$,
where $\alpha \in S^2$ is the direction of the propagation
of the incident wave.

We assume that the scatterers are small in the sense $ka<<1$.
Parameter $k>0$ is assumed fixed,
so the limits below are designated as limits $a\to 0$, and
condition $ka<<1$ is valid as $a\to 0$.

If $ka<<1$, then the following transformation of \eqref{e3} is
valid: \be\label{e4} u_M(x)=u_0(x)-\sum_{m=1}^M \frac
{e^{ik|x-x_m|}}{4\pi}A_m u_M(x_m) \int_{|y-x_m|<a}\frac
{dy}{|x-y|}[1+o(1)]. \ee To get \eqref{e4} we have used the folowing
estimates:
$$|x-x_m|-a\leq |x-y|\leq |x-x_m|+a, \qquad |y-x_m|\leq a.$$
These estimates imply
$$e^{ik |x-y|}=e^{ik |x-x_m|}[1+o(1)],$$
provided that $|y-x_m|\leq a$ and $a\to 0$.

We also have taken into account that, as $a\to 0$, one has:
$$\max_{x\in B_m}|u_M(x)-u_M(x_m)|=o(1).$$
A proof of this statement is given in \cite{R509}.

Here is a sketch of another proof of the above statement. Equation 
\eqref{e3} has a unique solution
because it is an equation with a compactly supported
bounded uniformly with respect to $M$ potential $p$. Its solution
satisfies a homogeneous Schr\"odinger equation and the radiation 
condition. Thus, this solution is unique. 
This solution is uniformly (with respect to $M$) continuous
in $D$.

Yet another argument, different from the ones, given above,
can be outlined as follows. The limiting funtion $u_e(x)$
is in $H^2(D)$ by the standard elliptic regularity results,
so it is continuous in $D$. The function $u_M$ converges to
$u_e$ uniformly. Therefore, it satisfies the above inequality
as  $M\to \infty$, or, equivalently, as  $a\to 0$.

The fourth proof of the same statement can be based on a result
from \cite{R563}. Namely, a 
justification of a collocation method for solving
equation (17) (see below) for the limiting field $u_e$
is given  in \cite{R563}. From the arguments, given
there, one obtains the uniform convergence of
$u_M$ to $u_e$:
$$\lim_{M\to \infty}\max_{x\in D}|u_M(x)-u_e(x)|=0.$$

Let us return to equation  \eqref{e4}.
 
We want to prove that the sum in \eqref{e4} has a limit as $a\to 0$, and
to calculate this limit assuming that the distribution of small
inhomogeneities or, equivalently, the points $x_m$, is given by 
the following formula:
\be\label{e5}
\mathcal{N}(\Delta)=|V(a)|^{-1}\int_{\Delta}N(x)dx [1+o(1)]\qquad a\to 0,
\ee 
where
$V(a)=4\pi a^3/3$, $N(x)\geq 0$ is an arbitrary given function, continuous 
in the closure of $D$,  $0\leq N(x)<\bf{p}<1$, $\bf{p}$ is a constant 
arising in 
the problem of spheres packing,  and $\mathcal{N}(\Delta)$ is 
the number of 
small inhomogeneities in an arbitrary open subset $\Delta\subset D$. 

 The maximal value of the constant constant  $\bf{p}$
is the maximal ratio of the total volume of the packed spheres divided
by $|\Delta|$, where $|\Delta|$ is the volume of $\Delta$. 
The total volume of the balls, embedded in the domain  $\Delta$,
is equal to 
$$V_{\Delta}:=V(a)\mathcal{N}(\Delta)=\int_{\Delta}N(x)dx [1+o(1)].$$
If $\diam (\Delta)$ is sufficiently small and $x\in \Delta$ is
some point in $\Delta$, then 
$$V_{\Delta}=N(x)|\Delta|[1+o(1)]<\bf{p} |\Delta|.$$
Since the 
domain $\Delta$ is arbitrary, one lets $a\to 0$ and  concludes
that 
$$N(x)\leq \bf{p}<1.$$
{\it It is conjectured that maximal value of $\bf{p}$ is 
equal approximately $ 0.74$.} 
(see, e.g., \cite{S}).

There is a large literature on optimal packing of spheres (see, e.g.,
\cite{CS}, \cite{S}). For us the maximal value of $\bf{p}$ is not
important. What is important is the following conclusion: 

{\it One can
choose $N(x)\geq 0$ as small as one wishes, and still create any desired
potential $q(x)$ by choosing suitable $A(x)>0$}.

The 
total number  $M=\mathcal{N}(D)$
of the embedded small inhomogeneities is $M=O(\frac 1 {a^3})$, according
to \eqref{e5}.

Our basic new tool is the following lemma.

{\bf Lemma 1.} {\it If the points $x_m$ are distributed in a bounded
domain $D\subset \R^3$ according to \eqref{e5},
and $f(x)$ is an arbitrary Riemann-integrable in $D$
function, then the following limit exists:
\be\label{e6}
 \lim_{a\to 0}\sum_{m=1}^M f(x_m)V(a)=\int_Df(x)N(x)dx.
\ee
}

The class of Riemann-integrable functions is
precisely the class of  bounded almost everywhere
continuous functions, i.e., bounded functions with the set of
discontinuities of Lebesgue measure zero in $\R^3$.
The class of such functions  is
precisely the class of functions for which the Riemannian sums
converge to the integrals of the functions.

The result in \eqref{e6} we generalize to the class of functions
for which the integral in the right-hand side of \eqref{e6} 
exists as an improper integral.
{\it In this case $f(x)$ may be unbounded at some points $y$}, but
the limit $\lim_{\delta \to 0} \int_{D_\delta}f(x)N(x)dx$
exists and
$$\lim_{\delta \to 0} \int_{D_\delta}f(x)N(x)dx:=\int_Df(x)N(x)dx,$$
where $D_\delta:=D\setminus B(y,\delta)$, and $B(y,\delta)$ is the ball
centered at $y\in D$ and of radius $\delta$.
In this case the sum in \eqref{e6} is defined as follows:
\be\label{e6'}\lim_{a\to 0}\sum_{m=1}^M f(x_m)V(a):=\lim_{\delta \to 
0}\lim_{a\to
0}\sum_{x_m\in D_\delta} f(x_m)V(a).
\ee

The same definition is valid for the conclusion of Theorem 1,
which is our result.

{\bf Theorem 1.} {\it If the small inhomogeneities are distributed so that
\eqref{e5} holds, and $q_m(x)=0$ if $x\not\in B_m$, $q_m(x)=A_m$
if $x\in B_m$,
where $B_m=\{x: |x-x_m|<a\}$, $A_m:=A(x_m)$, and
$A(x)$ is a given in $D$ function such that the function
$q(x):=A(x)N(x)$ is Riemann-integrable,
then the limit
\be\label{e7}
 \lim_{a\to 0}u_M(x)=u_e(x)
\ee
does exist and solves problem \eqref{eq1'}-\eqref{eq2'} with the
potential
\be\label{e8}
q(x)=A(x)N(x).
\ee
}

There is a large literature on wave scattering by small inhomogeneities. A
recent paper, which we use, is \cite{R564}.
Some of the ideas of this
approach were earlier applied by the author to scattering by small
particles embedded in an inhomogeneous medium (see [3]-[23]).

 We apply Lemma 1 to the sum in \eqref{e4},
in which we choose $A_m:=A(x_m)$, where $A(x)$
is an arbitrary continuous in $D$ function
which we may choose as we wish. A simple calculation
yields the following formulas:
\be\label{e14}
\int_{|y-x_m|<a}|x-y|^{-1}dy=V(a)|x-x_m|^{-1}, \qquad |x-x_m|\geq a,
\ee
and
\be\label{e15}
\int_{|y-x_m|<a}|x-y|^{-1}dy=2\pi(a^2- \frac{|x-x_m|^2}{3}), \qquad
|x-x_m|\leq a.
\ee
Therefore, the sum in  \eqref{e4} is of the form  \eqref{e6} with
\be\label{e15'}f(x_m)=\frac 
{e^{ik|x-x_m|}}{4\pi|x-x_m|}A(x_m)u_M(x_m)[1+o(1)].
\ee
Applying Lemma 1,
one concludes that the limit $u_e(x)$ in  \eqref{e7} does exist and solves
the
integral equation
\be\label{e16}
u_e(x)=u_0(x)-\int_D \frac {e^{ik|x-y|}}{4\pi|x-y|}q(y)u_e(y)dy,
\ee
where $q(x)$ is defined by formula \eqref{e8}.

Applying the operator $\nabla^2 +k^2$ to  \eqref{e16}, one verifies that
the function  $u_e(x)$ solves problem \eqref{eq1'}-\eqref{eq2'}.

Thus, the conclusion of Theorem 1 is established. 

In \eqref{e15'}  $f(x_m)$ depends on $a$ through $u_M(x_m)$.
Nevertheless Lemma 1 is applicable because $u_M(x)$ converges
to the solution $u_e(x)$ of eqution \eqref{e16} as $a\to 0$.
This follows from the results in \cite{R563} and \cite{R509}.
These results give two independent ways to prove convergence 
of $u_M$ to $u_e$ as $a\to 0$. The results in \cite{R509}
yield compactness of the set $\{u_M(x)\}$, while the results in
  \cite{R563} convergence of a version of  the collocation 
method for solving equation  \eqref{e16} is proved. This version of the
 collocation method leads to solving the following linear algebraic system 
\be\label{elas}
u_j=u_{0j}-\sum^n_{m=1, m\neq j}g_{jm}q_mu_m V(a), \quad 1\leq j \leq m,
\ee  
where $u_j:=u(x_j), \, q_m:=q(x_m), $ and 
$g_{jm}:=\frac {e^{ik|x_j-x_m|}}{4\pi|x_j-x_m|}$.
It is proved in  \cite{R563}
that the linear algebraic system \eqref{elas} has a solution and
this solution is unique for all sufficiently small $a$,
and the function 
$$u^{(n)}(x):=\sum^n_{m=1} u_j \chi_j(x)$$
converges to $u_e(x)$ uniformly:
$$\lim_{n\to \infty}||u_e(x)-u^{(n)}(x)||_{C(D)}=0,$$
where $ \chi_j(x)=1 \quad x\in D$,  $ \chi_j(x)=0 \quad x\not\in D$.

In particular, it is proved in \cite{R563} that a collocation 
method for solving equation \eqref{e16} leads to sums, similar to
the sum on the left-hand side of \eqref{e6}, and the collocation method,
studied in \cite{R563},  converges to the unique solution of the 
equation \eqref{e16}. This means that the function $u_M(x)$
converges to  $u_e(x)$.
 
{\it Remark 1.}
The quantities $A_m$ in the definition of $q_m$ can be {it tensors}. In 
this
case {\it the refraction coefficient $n^2(x)$ of the created material is
a tensor, and $u$ is a vector field}.

{\it Remark 2.} The recipe for creating materials with the desired
refraction coefficient by the method, based on Theorem 1, can be
summarized as follows:

{\it Recipe 1:  Given a bounded domain filled with the material with the
refraction
coefficient $n_0^2(x)$, one embeds the  small inhomogeneities
$q_m$, $q_m=0$ for $|x-x_m|>a$, $q_m=A_m:=A(x_m)$, where
the points $x_m$ are distributed by the rule  \eqref{e5},
and one chooses $A(x)$ and $N(x)\geq 0$,  so that 
$A(x)N(x)=q(x)$,
where $q(x)$ is the desired potential, or which is the same,
the desired refraction coefficient $n^2(x):=1-k^{-2}q(x)$. 
The material, created by the embedding
of these small inhomogeneities, in the limit $a\to 0$,
has the refraction coefficient $n^2(x)=1-k^{-2}q(x)$.

One can fix $N(x)\geq 0$ and choose $A(x)$ such that
the function $q(x)=A(x)N(x)$ is a desired function, so that
$n^2(x)=1-k^{-2}q(x)$ is a desired function.}

We call this recipe {\it Recipe i}, where {\it i} stands for 
inhomogeneities.

\section{A discussion of two recipes}

In \cite{R552} a different recipe was proposed for creating materials with 
a desired refraction coefficient. We call the recipe from \cite{R552}
{\it Recipe p}, where {\it p} stands for particles. {\it Recipe p} 
consists of 
embedding small particles, for example,
balls of radius $a$, $ka<<1$, into a given material according to the
distribution law, similar to the one in \eqref{e5}, but with
$a^{2-\kappa}$
in place of $V(a)$, where $\kappa\in (0,1]$ is a parameter experimenter
can choose at will.  The physical properties of the embedded particles are
described by the boundary impedance
$\zeta_m=\frac{h(x_m)}{a^{\varkappa}}$. The function $h(x)$, determining
the boundary impedances, can also be chosen as one wishes. We
refer the reader
to the paper \cite{R552} for details. Numerical results on the 
implementation of {\it Recipe p} are given in \cite{R580} and 
\cite{R583}.

There are two technological problems that should be solved in order that
{{\it Recipe p} be implemented practically.

{\bf Technological Problem 1:} {\it How does one embed many small 
particles in a given
material
so that the desired distribution law
is satisfied?}

{\bf Technological Problem 2:} {\it a) How does one prepare a small 
particle with the desired boundary
impedance $\zeta_m = \frac{h(x_m)}{a^{\varkappa}}$?

b) How can one prepare a small particle with the desired
dispersion of the boundary impedance, that is, the desired 
$\omega$-dependence of $h=h(x,\omega)$?}

The motivation for the last question 
is the following: by creating material with the refraction coefficient
depending on the frequency $\omega$ in a desired way, one can, for
instance,
prepare materials with negative refraction.

The first problem, possibly, can be solved by {\it stereolitography}.

The second problem one
should be able to solve because the limiting cases
$\zeta_m=0$ (hard particles) and
$\zeta_m=\infty$
(soft particles) can be prepared, so that any intermediate value
of $\zeta_m$ one should be able to
prepare as well. 

The author formulates the above technological problems
in the hope that
engineers get interested and solve them practically.

In {\it Recipe p} the number $M$ of the embedded particles is
$M:=M_2=O(\frac 1{a^{2-\kappa}})$, while in {\it Recipe i} the number $M$
is $M:=M_1=O(\frac 1{a^3})$. Thus, $M_1>>M_2$ as $a\to 0$. So, in
recipe 2 one has to embed much smaller number of small particles
than in {\it Recipe i}. This is an advantage of {\it Recipe p}
over {\it Recipe i}. The disadvantage
of {\it Recipe p}, compared with {\it Recipe i}, is in the 
possible practical difficulties of creating
particles with the prescribed very large boundary impedances
$\zeta_m=O(\frac 1 {a^\kappa})$ when $a$ is very small.

\subsection{Negative refraction}

Material with negative refraction is, by definition, a material in which
group velocity is directed opposite to the phase velocity, (see \cite{A}
and references therein). Group velocity is defined by the formula
$\mathbf{v}_g = \nabla_\mathbf{k}\omega(\mathbf{k})$.  Phase velocity
$\mathbf{v}_p$ is directed along the wave vector
$\mathbf{k}^0=\frac{\mathbf{k}}{|\mathbf{k}|}$. In an isotropic material
$\omega = \omega(\mathbf{|k|})$, and 
$\omega n(x,\omega)=c|\mathbf{k}|$. 
Differentiating this equation yields \begin{equation}
\label{eq29} \nabla_{\mathbf{k}}\omega \big{[}n(x,\omega)+ \omega
\frac{\partial n}{\partial \omega }\big{]} = c\mathbf{k}^0. \end{equation}
Thus, \begin{equation} \label{eq30} \mathbf{v}_g \big{[}n+\omega
\frac{\partial n}{\partial \omega }\big{]} = c\mathbf{k}^0. \end{equation}
Wave speed in the material is
$|\mathbf{v}_p|=\frac{c}{n(x,\omega )}$, where $c$ is the wave speed in
vacuum (in $D'$), and $n(x,\omega )$ is a scalar, and $\mathbf{v}_p$ is 
directed along $\mathbf{k}^0$.

For $\mathbf{v}_g$ to be directed opposite to $\mathbf{v}_p$ 
it is necessary and sufficient that
\begin{equation}
\label{eq31}
n+\omega \frac{\partial n}{\partial \omega } < 0.
\end{equation}
If the new material has the function $n(x,\omega)$ satisfying 
\eqref{eq31}, then the new
material has negative refraction.

One can create material with the refraction coefficient $n^2(x,\omega )$
having a desired dispersion, i.e., a desired frequency dependence, by
choosing function $h=h(x,\omega )$ properly.

Let us formulate the technological problem solving of which allows one to
implement practically {\it Recipe p} for creating materials with negative
refraction.

{\bf Technological Problem 3:} {\it How does one prepare a small particle 
with the desired frequency dependence of the boundary impedance
$\zeta_m=\frac{h(x_m,\, \omega)}{a^{\kappa}}$, where $h(x,\omega )$
is a given function?}

This problem is the same as {\it Technological Problem 2b)}, formulated 
above. 

\subsection{Wave-focusing property}

Let us formulate the problem of preparing a material with a
desired wave-focusing property. The refraction coefficient
is related to the potential by the formula 
$$q(x)=k^2-k^2n^2(x).$$
We assume in this Section that $k>0$ is fixed, and $\alpha\in S^2$,
the incident direction (that is, the direction $\alpha$ of the incident
plane wave
$e^{ik\alpha \cdot x}$) is also fixed.

{\it Problem: Given an arbitrary fixed $f(\beta) \in L^2(S^2)$, an
arbitrary small fixed
$\epsilon>0$, an arbitrary fixed $k>0,$ and an arbitrary fixed
$\alpha\in S^2$, can one
find $q\in L^2(D)$, $q=0$ in $D'=R^3\backslash D$, such that
\begin{equation}
        \| A_q(\beta) - f(\beta)\|_{L^2(S^2)} <\epsilon,
\end{equation}
where $A_q(\beta):=A_q(\beta,\alpha,k)$ is the scattering amplitude,
corresponding to $q$, at fixed $\alpha$ and $k$, and $q:=k^2-k^2
n^2(x)$?}

The answer is {\it yes}, and an algorithm for finding such a $q$,
and, therefore, such an $n^2(x)$, is given in \cite{R523}.
See also \cite{R524}, \cite{R532}. The above problem is the inverse 
scattering problem with the scattering data  given at a fixed $k$ and 
a fixed direction  $\alpha$ of the incident plane wave. 

{\it There are many potentials $q$ which solve the above problem.} It is
possible to choose from the set of these potentials the one
which satisfies some additional properties, for example, one can choose
$q$ to be arbitrarily smooth in $D$, one can try to choose $q$ with 
non-negative imaginary part, etc.

The desired radiation pattern $f(\beta)$ can, for example, be
equal to $1$ in a given solid angle and equal to $0$ outside this solid
angle. In this case the scattered field is scattered
predominantly into the desired solid angle.

This is why we call such a material a material with wave-focusing 
property.

\end{document}